\documentstyle[twocolumn,epsf]{jpsj}

\def\runtitle{Nuclear Spin Relaxation in Hole Doped Two Leg Ladders}
\def\runauthor{Shigeki {\sc Fujiyama}, Masashi {\sc Takigawa}, Naoki {\sc
Motoyama}, Hiroshi {\sc Eisaki} and Shin-ichi {\sc Uchida}}

\title
{ Nuclear Spin Relaxation in Hole Doped Two-Leg Ladders }

\author
{ Shigeki {\sc Fujiyama}, Masashi {\sc Takigawa}, Naoki {\sc
Motoyama}$^{1}$,
\\ Hiroshi {\sc Eisaki}$^{1}$ and Shin-ichi {\sc Uchida}$^{1}$
}

\inst
{The Institute for Solid State Physics, The University of Tokyo, Roppongi,
Minato-ku, Tokyo 106-8666 \\ The Department of Advanced Materials Science,
The University of Tokyo, Yayoi, Bunkyo-ku,Tokyo 113-8656 $^{1}$}
\recdate
{March 7, 2000}

\abst
{The nuclear spin-lattice relaxation rate ($1/T_{1}$) has been measured in
the single crystals of hole doped two-leg ladder compounds
Sr$_{14-x}$Ca$_{x}$Cu$_{24}$O$_{41}$ and in the undoped parent material
La$_6$Ca$_8$Cu$_{24}$O$_{41}$. Comparison of $1/T_{1}$ at the Cu and the
two distinct oxygen sites revealed that the major spectral weight of low
frequency spin fluctuations is located near $q \sim (\pi, \pi)$ for most of
the temperature and doping ranges investigated. Remarkable difference in
the temperature dependence of $1/T_1$ for the two oxygen sites in the
heavily doped $x$=12 sample revealed reduction of singlet correlations
between two legs in place of growing antiferromagnetic correlations along
the leg direction with increasing temperature. Such behavior is most likely
caused by the dissociation of bound hole pairs. }

\kword
{ spin ladder, spin gap, hole doping, NMR }

\begin{document}
\sloppy
\maketitle

There has been considerable interest in the hole-doped two-leg spin ladders
since superconductivity was predicted to occur in such
systems.~\cite{Dagotto1996} It has been well established that undoped
two-leg spin ladders have a resonating valence bond ground state and a
finite gap for spin excitations. How the spin and charge dynamics as well
as the nature of the ground state evolve with hole-doping is an issue of
current interest. Both analytic~\cite{Sigrist1994,Kishine1998} and
numerical~\cite{Hayward1996,Troyer1996} theories predict that doped holes
would be bound in pairs with a finite spin gap, leading to $d$-wave
superconductivity unless dominated by charge order (CDW) instability. This
prediction stimulated many efforts to realize hole doping into two-leg
ladders. Of these, Sr$_{14-x}$Ca$_{x}$Cu$_{24}$O$_{41}$ is the only
hole-doped material with two-leg ladder structure so far known that shows
superconductivity.~\cite{Uehara1996}

The structure of this compound consists of alternating stack of the
Cu$_{2}$O$_{3}$ two-leg ladder layers shown in Fig.~\ref{fig:CuT1} (a) and
the CuO$_{2}$ one dimensional chain layers with the Sr layers between
them.~\cite{McCarron1988} The average valence of Cu is 2.25, i.e. 1/4 holes
per Cu are distributed among the ladder and the chain layers. Holes in the
chain layers are highly localized and do not contribute to charge
transport. The hole density in the ladder layers ($P_L$) of the $x$=0
sample (Sr$_{14}$Cu$_{24}$O$_{41}$) is estimated to be rather small ($P_L
\sim 0.05$) from optical measurements~\cite{Osafune1997}. Substitution with
Ca for Sr promotes transfer of holes from the chain layers to the ladder
layers. Thus the electrical resistivity decreases with $x$ and
superconductivity appears for $x \ge 11$ under high pressure above 3
GPa.~\cite{Uehara1996,Nagata1998} By substituting Sr or Ca with La, total
number of holes can be reduced. All Cu ions are divalent carrying spin 1/2
in La$_{6}$Ca$_8$Cu$_{24}$O$_{41}$ (hereafter abbreviated as La6Ca8), which
can be regarded as the undoped parent material.

Spin excitations in the ladder layers have been studied by neutron
scattering and NMR. Neutron scattering experiments in undoped La6Ca8 have
shown a finite gap $\Delta$=32meV for spin excitations at the wave vector
$(\pi, \pi)$.~\cite{Matsuda1999} The gap persists upon doping and its
magnitude remains nearly independent of doping up to
$x$=12.~\cite{Eccleston1998,Regnault1999,Katano1999} Measurements of the
nuclear spin-lattice relaxation rate ($1/T_{1}$) and the NMR frequency
shift at the ladder Cu sites also show approximately activated temperature
dependences, apparently indicating a finite spin
gap.~\cite{Tsuji1996,Carretta1997,Takigawa1998,Magishi1998,Imai1998}
However, it has been puzzling that the activation energy decreases
dramatically with increasing $P_{L}$ contrary to the neutron results.
Recent results of Cu NMR for $x$=12 under high
pressure~\cite{Mayaffre1998,Piskunov2000} indicate that the spin gap
vanishes in the normal state of the superconducting materials. Thus it is
still controversial whether the superconductivity emerges from preformed
hole pairs in the RVB spin singlet background as the theories originally
predicted.


\begin{fullfigure}
    \epsfxsize=17cm
\centerline{\epsfbox{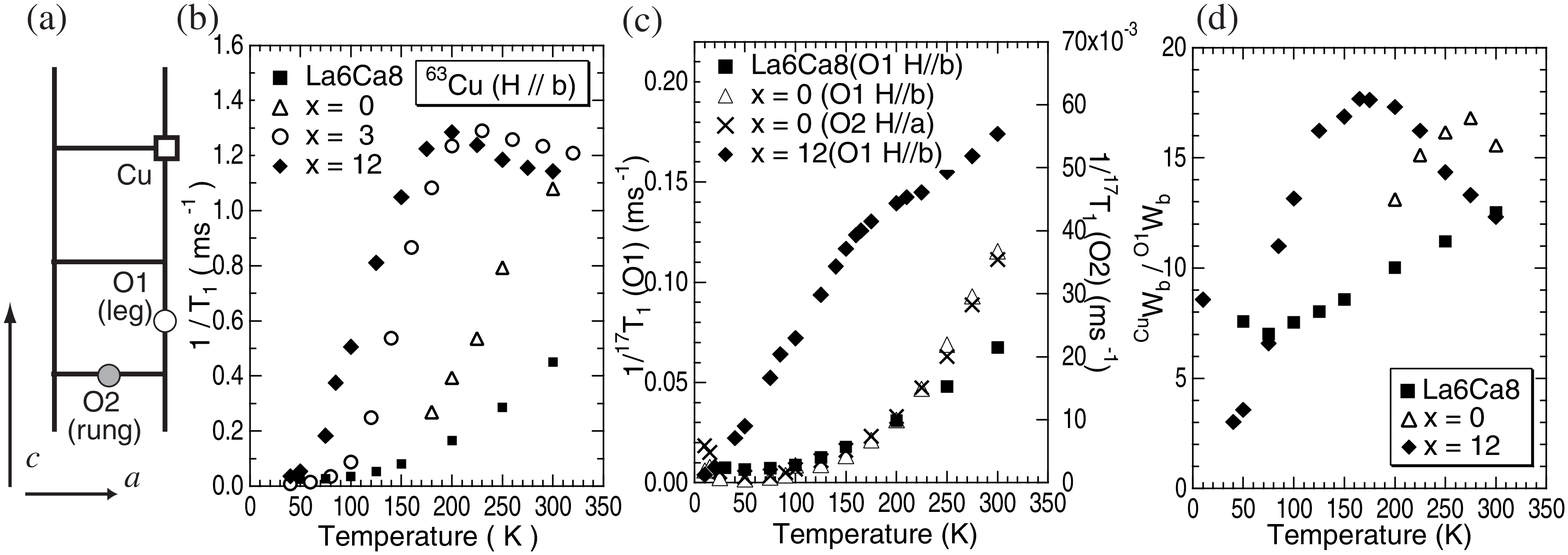}}
\caption{(a) Structure of the ladder layers. (b) Temperature dependence of
$(1/^{\rm Cu}T_{1})_b$ for La6Ca8, $x=$0, 3 and $x$=12. (c) Temperature
dependence of $(1/^{\rm O1}T_{1})_b$ for La6Ca8, $x=0$, and $x=12$. The
results of $(1/^{\rm O2}T_{1})_a$ for $x$=0 are also shown. (d) The
temperature dependence of the ratio $^{{\rm Cu}}W_{b}/^{{\rm O1}}W_{b}$.}
\label{fig:CuT1}
\end{fullfigure}
In this paper, we discuss low energy spin dynamics at ambient pressure
based on the results of $1/T_1$ measurements in three single crystals, the
undoped La6Ca8 and the hole-doped Sr$_{14-x}$Ca$_{x}$Cu$_{24}$O$_{41}$ with
$x$=0 and $x$=12. Since a two-leg spin ladder is a quasi one-dimensional
system with short range antiferromagnetic correlations, low energy spin
excitations should be located near $q_{x}$=0 or $\pi$ (2$k_F$) in the
reciprocal space, where $q_{x}$ is the wave vector along the leg direction.
Spin excitations are further specified by the symmetry with respect to the
interchange of two legs. Symmetric (antisymmetric) excitations are
associated with $q_{y}$=0 ($q_{y}$=$\pi$). Thus there are four important
spots in the $q$-space that could contribute to the low energy spin
fluctuations. All of them contribute to $1/T_{1}$ at the Cu sites. However,
antiferromagnetic spin fluctuations along the leg direction with $q_x \sim
\pi$ do not contribute to $1/T_1$ at the leg oxygen (O1) sites, neither do
fluctuations with $q_y \sim \pi$ to $1/T_1$ at the rung oxygen (O2) sites.
Therefore, measurements of $1/T_1$ at different sites allow us to
distinguish low energy spin fluctuations at different wave vectors.

Our results indicate that the dominant weight of low frequency spin
fluctuations is located near $(\pi, \pi)$ for most of the doping and
temperature range investigated. This implies that the activation energy of
$1/T_1$ at the Cu sites is determined by the damping of excitations near
$(\pi, \pi)$ and should not be identified as the spin gap. The discrepancy
with the neutron results is thus resolved. The mechanism of damping,
however, depends on doping and is closely correlated with the transport
behavior.

The single crystals used in the NMR experiments were grown in oxygen
atmosphere by the travelling solvent floating zone
method.~\cite{Motoyama1997} The crystals were then annealed in oxygen gas
containing 45\% $^{17}$O for isotopic exchange. The temperature dependence
of $1/T_{1}$ at the Cu and O1 sites are shown in Fig.~\ref{fig:CuT1} (b)
and (c), respectively. A general expression for $1/T_{1}$ at $k$-th sites
is given in terms of dynamic spin correlation function $S({\bf q},
\omega_{n})$ as $(1/^{k}T_{1})_{\alpha} = 2(^{k}\gamma_{n})^{2} \sum_{q}
(^{k}~F_{\beta}({\bf q})^{2}+^{k}~F_{\gamma}({\bf q})^{2}) S({\bf q},
\omega_{n})$.
Here, $\alpha$ is the direction of the external field, $\beta$ and $\gamma$
denote the directions perpendicular to $\alpha$, $^{k}\gamma_{n}$ is the
gyromagnetic ratio of the $k$-th nuclei, $\omega_n$ is the nuclear
resonance frequency (several tens of MHz), and $^{k}F_{\alpha}({\bf q})$ is
the wave-vector dependent hyperfine coupling constant. The ${\bf q}$
dependence of $F({\bf q})$ describes the filtering of spin fluctuations in
the $q$-space for each site mentioned before. The values of $F_{\alpha}(0)$
have been determined from the NMR frequency shift measurements up to an
overall multiplicative factor.~\cite{Fujiyama} We define the normalized
$1/T_1$ by
\begin{equation}
^{k}W_{\alpha} \equiv C \frac{(1/^{k}T_{1})_{\alpha}}
{(^{k}\gamma_{n})^{2}(^{k}F_{\beta}(0)^{2}+^{k}F_{\gamma}(0)^{2})} ,
\label{eq:normt1}
\end{equation}
where $C = (^{\rm Cu}\gamma_{n})^{2}(^{\rm Cu}F_{a}(0)^{2}+^{\rm
Cu}F_{c}(0) ^{2})$ so that $^{\rm Cu}W_{b}$ is identical to $(1/^{\rm
Cu}T_{1})_{b}$. If there is no filtering, the value of $^{k}W_{\alpha}$
should be identical for all nuclei. Experimentally, however, the values and
temperature dependence of $^{k}W_{\alpha}$ are quite different for the Cu
and O1 sites, as one can clearly see by plotting the ratio $^{\rm
Cu}W_{b}/^{\rm O1}W_{b}$ shown in Fig.~\ref{fig:CuT1} (d).

{\em 1. Undoped La6Ca8}: In the undoped La6Ca8, $1/T_1$ shows an activated
temperature dependence above 150 K with the activation energy of about 
540 K
at both sites. Below about 100 K, however, $1/T_1$ shows a $T$-independent
finite value. This is most likely due to the coupling to the magnetic
CuO$_2$ chain layers. It has been established that for one dimensional spin
systems with a finite gap, spin correlation functions near zero frequency
at low temperatures are described by the two magnon processes of thermally
excited magnons with the momentum transfer ${\bf q} \sim (0,
0)$.~\cite{Troyer1994,Jolicoeur1994} Then we expect the normalized $1/T_1$
to be identical for all sites. However, $^{\rm Cu}W_{b}/^{\rm O1}W_{b}$
shown in Fig.~\ref{fig:CuT1} (d) is much larger than one for the whole
temperature range and increases with temperature above 150 K. This indicates
that $S(\pi, \pi) \gg S(0, 0)$ and $S(\pi, \pi)$ is more enhanced at higher
temperatures. (For the rest of the paper, we omit $\omega_{n}$ in $S({\bf
q}, \omega_{n})$ so that $S(q_x, q_y)$ represents $S(q_x, q_y, \omega=0)$.)

The magnitude of $S(\pi, \pi)$ is given by the $\omega$=0 spectral weight
of the magnon states broadened by life time (damping) effects due to
multimagnon processes such as scattering of single magnons with two magnon
continuum.~\cite{Jolicoeur1994} It has the temperature dependence
$\exp(-2\Delta/T)$ in the low temperature limit and, therefore, should be
negligible compared with $S(0, 0) \sim \exp(-\Delta/T)$. However, numerical
studies have shown that the validity of such argument is limited to very
low temperatures. Monte Carlo~\cite{Sandvik1996} and DMRG~\cite{Naef1999}
calculations have shown that ($\pi,\pi$) contribution becomes dominant even
the temperature is as small as 0.3$\Delta$ when the exchange along the leg
is equal or larger than the exchange along the rung. ~\cite{Ivanov1999}
Such anisotropy of the exchange constants is indeed suggested from analysis
of the NMR frequency shift at the two oxygen sites
.~\cite{Imai1998,Fujiyama}

Imai {\it et al}. found a crossover near 430 K in La6Ca8, above which
$1/^{\rm Cu}T_1$ and $1/^{\rm O1,O2}T_1$ show remarkably different
temperature dependences.~\cite{Imai1998} They argued that two magnon
processes with ${\bf q} \sim (0, 0)$ is dominant below this temperature.
However, the comparison of the magnitude of $1/^{\rm Cu}T_1$ and $1/^{\rm
O1}T_1$ in our data and closer examination of their temperature
dependences indicate that ${\bf q} \sim (\pi, \pi)$ contribution becomes
dominant already for $T \ge 150$K.

{\em 2. Lightly doped x=0}: The lightly doped $x$=0 sample shows
semiconducting temperature dependence of resistivity although it is
reasonably conducting along the leg direction (5m$\Omega$cm) at room
temperature. Comparison of the $1/T_1$ data for La6Ca8 and $x=0$ in
Figs.~\ref{fig:CuT1} (b) and (c) shows that slight hole-doping causes
significant enhancement of $1/T_1$ above 200 K, where holes become mobile
and responsible for incoherent transport. This enhancement is much larger
at the Cu sites than at the O1 sites.

Fitting of the $1/T_1$ data above 200 K to an activated temperature
dependence results in quite a large value of the activation energy ($\sim$
800 K) at both sites, which is much larger than the spin-gap of 32.5 meV
obtained from neutron scattering measurements at 20 K~\cite{Eccleston1998}.
The value of $^{\rm Cu}W_{b}/^{\rm O1}W_{b}$ for $x$=0 is larger than that
for La6Ca8 (Fig.~\ref{fig:CuT1} (d)), indicating that $S(\pi, \pi)$ is
further enhanced over $S(0, 0)$ by hole-doping. This result implies that
incoherent motion of holes further accelerates the magnon damping, thereby
increases the zero energy spectral weight near $(\pi, \pi)$ and makes
$S(\pi, \pi)$ by far the most dominant contribution to $1/^{\rm Cu }T_1$.
The two oxygen sites show similar temperature dependence of $1/T_1$, as
shown in Fig.~\ref{fig:CuT1} (c).

Several anomalies of charge properties have been observed near 200 K for
$x$=0, supporting our argument relating the enhanced $1/T_1$ above this
temperature with motion of holes. At the Cu sites, $1/T_1$ measured by zero
field NQR is strongly enhanced below 180 K by quadrupolar relaxation due to
slow fluctuations of electric field gradient.~\cite{Takigawa1998} Analysis
based on the standard motional narrowing theory indicates that glassy
charge freezing occurs in this temperature range~\cite{Takigawa1998}. The
electrical resistivity plotted against $1/T$ also shows clear change of the
activation energy from 1400 K for $T \le 180$K to 2200 K for $T \ge
180$K.~\cite{Motoyama1997} The values of electric field gradient at the Cu
and two oxygen sites also show sudden steep increase above
200 K.~\cite{Imai1998,Takigawa1998} These results indicate that both static
charge distribution and charge dynamics change their behavior near 200 K.

\begin{figure}[tbp]
\centering
\epsfxsize=8cm
\centerline{\epsfbox{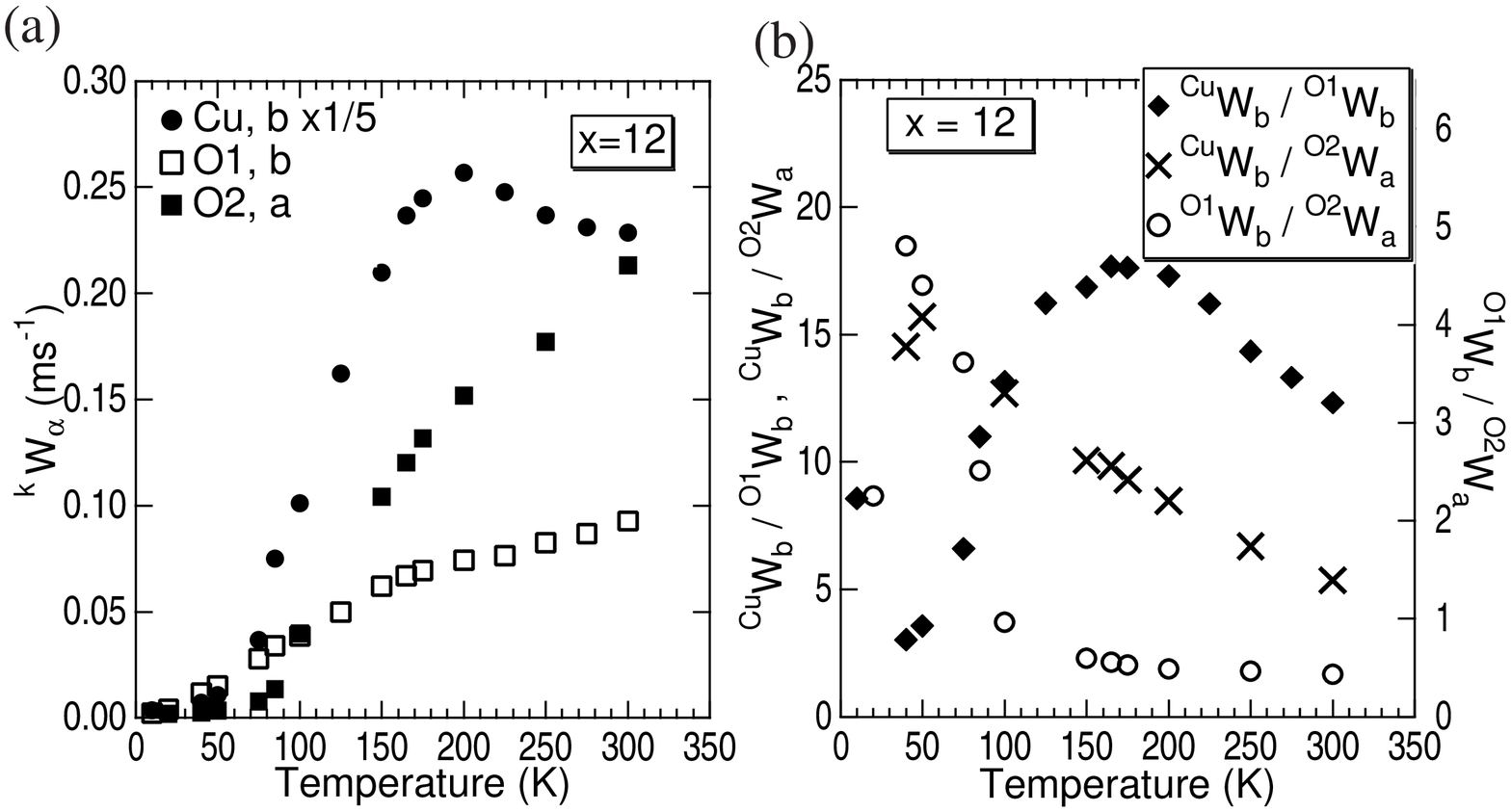}}
\caption{(a)The normalized relaxation rates defined by eq.\ref{eq:normt1}
at the Cu and the two oxygen sites in $x=12$. (b) The ratios of the
normalized relaxation rates for $x$=12.}
\label{fig:Ca12T1}
\end{figure}
{\em 3. Heavily doped $x=12$}: The data for $x$=12 sample in
Fig.~\ref{fig:CuT1} (b) and (c) indicate that further hole-doping up to
$P_L$=0.2 ($x$=12) significantly lower the temperature range in which
$1/T_1$ shows rapid variation. In order to compare $1/T_1$ at various sites
more clearly, we plotted the normalized relaxation rates $^{\rm Cu}W_{b}$,
$^{\rm O1}W_{b}$, and $^{\rm O2}W_{a}$ in Fig.~\ref{fig:Ca12T1} (a). Their
ratios are plotted in Fig.~\ref{fig:Ca12T1} (b).

Below about 60 K, $^{\rm Cu}W_{b}/^{\rm O1}W_{b}$ is relatively small
($\sim$ 2), indicating absence of strong enhancement of $S(\pi, \pi)$
compared with other region in ${\bf q}$-space. In the intermediate
temperature range from 60 K to 180 K, where rapid increase of $1/^{\rm
Cu}T_1$ is observed, $^{\rm Cu}W_{b}/^{\rm O1}W_{b}$ also increases steeply
and shows a broad maximum near 180 K. This means that $S(\pi, \pi)$ is
getting dominant over other components. Therefore, the variation of
$1/^{\rm Cu}T_1$ in this temperature range with an apparent small
activation energy ($\sim 200$ K) is not associated with thermal excitations
across the gap but caused by accumulation of low frequency spectral weight
near $(\pi, \pi)$. Thus it is not in conflict with the neutron results that
shows a large spin gap of 370 K in the low temperature limit ($T $=7 K).

A remarkable feature for the heavily doped material is that the two oxygen
sites show quite different temperature dependences of $1/T_1$. Below 
60 K,
the ratio $^{\rm O1}W_{b}/^{\rm O2}W_{a}$ is large ($\sim 5$), indicating
that $S(0, \pi) \geq S( \pi, 0), S(0, 0)$. As the temperature increases
$^{\rm O1}W_{b}/^{\rm O2}W_{a}$ falls rapidly and becomes smaller than one
above 100 K, indicating that $S(\pi, 0) \geq S(0, \pi), S(0, 0)$. The
temperature dependence of $^{\rm O1}W_{b}/^{\rm O2}W_{a}$ combined with
that of $^{\rm Cu}W_{b}/^{\rm O1}W_{b}$ suggests that singlet correlations
along the rung becomes weaker with increasing temperature in place of the
growing antiferromagnetic correlations along the leg. We should recall that
such behavior has not been seen in $x$=0 sample, where the two oxygen sites
behave similarly up to room temperature. The spin susceptibility deduced
from the shift measurements also shows rapid increase above 60 K and
saturation above 180 K.~\cite{Magishi1998,Fujiyama}

Above 180 K the behavior of $1/T_1$ is quite different for all three sites.
While $1/^{\rm Cu}T_1$ shows gradual decrease after passing through a broad
maximum around 180 K, $1/^{\rm O1}T_1$ and $1/^{\rm O2}T_1$ continue to
increase. While $1/^{\rm O1}T_1$ varies almost linear in $T$, $1/^{\rm
O2}T_1$ increases more steeply, implying rapid enhancement of $S(\pi, 0)$.
This is correlated with the continuous decrease of $^{\rm Cu}W_{b}/^{\rm
O2}W_{a}$ to a modest value ($\sim 4$), indicating that $S(\pi, 0)$ is
getting comparable to $S(\pi, \pi)$ at high temperatures. All these results
suggest that the two legs of the ladder unit become gradually decoupled
magnetically, hence no distinction between $q_y$=0 and $q_y$=$\pi$, and the
spin gap near $q_x \sim \pi$ is being filled with low frequency spin
fluctuations. Indeed nearly constant $1/^{\rm Cu}T_1$ and approximately
$T$-linear $1/^{\rm O1}T_1$ are what would be expected for an isolated spin
chains with gapless antiferromagnetic spin
fluctuations.~\cite{Takigawa1996,Sachdev1994} These results are consistent
with the neutron scattering experiments for $x$=11.5 by Katano {\it et
al}.~\cite{Katano1999} At $T$=7 K the energy scan spectrum at $(\pi, \pi)$
shows a peak at 32.5meV with no weight near $\omega \sim 0$. However, the
spectrum at $T$=200 K shows substantial weight below 10meV, even though the
peak still appears at a high energy near 20 meV.

The temperature variation of the spin correlation discussed above is
closely correlated with the transport behavior.~\cite{Motoyama1997}
Although the resistivity along the leg direction shows nearly $T$-linear
metallic behavior down to 60 K, semiconducting behavior starts below 200 K
along the rung direction. Motoyama {\it et al}.~ \cite{Motoyama1997} argued
that such behavior is an evidence for the hole pairing as has been
predicted theoretically.~\cite{ Sigrist1994,Kishine1998,
Hayward1996,Troyer1996} When holes are bound in pairs, coherent motion of
hole pairs is possible only along the leg directions. Since the probability
of interladder pair hopping is small, current along the rung direction has
to be carried by quasiparticles that result from thermal dissociation of
bound hole pairs. This picture is supported also by recent optical
measurements.~\cite{Osafune1999}

Our results appear to be consistent with this argument. Starting from the
low temperature side, dissociation of hole pairs begins to occur at about
60 K and almost completes near 180 K. Since hole pairing should be a
necessary condition for the persistence of spin gap when doped with holes,
dissociated quasiparticles carrying spin 1/2 will create spectral weight
within the magnon gap. Moreover, such mobile quasiparticles will cause
strong damping of magnons, filling the spin gap near $(\pi,
\pi)$. It is also physically plausible that singlet correlations between
the two legs is disturbed by motion of such quasiparticles, accounting for
different behavior at the O1 and O2 sites. When hole pairs are completely
unbound, the two legs will be highly decoupled and asymptotically behave as
gapless Tomonaga-Luttinger liquid with strong antiferromagnetic
correlation.

Numerical calculations for doped $t-J$ ladders have indeed demonstrated two
distinct types of spin excitations at $T$=0.~\cite{Troyer1996} One is
associated with the singlet-triplet excitations across the spin gap that
evolve continuously from the magnon excitations in undoped material. The
other is associated with the dissociation of bound hole pairs. Since the
latter type has smaller excitation energy, it is reasonable that this
process occurs at much lower temperatures than the spin gap in undoped
material.

Magishi {\it et al}. argued based on their data of $1/^{\rm Cu}T_1$ for
$x$=11.5 that the opening of spin gap indicated by sudden decrease of
$1/^{\rm Cu}T_1$ below 180 K coincides with the onset of hole pairing
inferred from the resistivity data.~\cite{Magishi1998} We should mention,
however, that such crossover in $1/^{\rm Cu}T_1$ is not limited to the
heavily doped materials but common to wide range of doping. For example,
Imai {\it et al}. observed similar crossover at 330 K (230 K) for $x$=0
($x$=3), for which no evidence of hole paring has been observed in
transport or optical measurements. This indicates that for small doping,
localization of individual holes rather than hole pairing is sufficient to
restore spin gap behavior at low temperatures. We consider that the
variation of singlet correlations along the rung is indicated by the
temperature dependence of the ratio $^{\rm O1}W_{b}/^{\rm O2}W_{a}$ is the
key factor related to the hole pairing and dissociation.

In conclusion, comparison of $1/T_1$ data at different sites indicates that
the dominant weight of low frequency spin fluctuations is located near
$(\pi, \pi)$ for most of the temperature and doping range we studied. The
activated temperature dependence of $1/T_1$, therefore, describes thermal
damping of magnons near $(\pi, \pi)$. The mechanism for damping depends on
doping. In undoped La6Ca8, multimagnon processes become important even when
the temperature is smaller than the gap, consistent with the recent DMRG
calculation.~\cite{Naef1999} In the lightly doped $x$=0 sample, motion of
holes above 200 K further accelerates the magnon damping. In the heavily
doped $x$=12 sample, we propose that dissociation of bound hole pairs
creates low energy spin fluctuations within the spin gap. Dissociated
quasiparticles moving along the leg reduce singlet correlations along the
rung in place of the growing antiferromagnetic correlations along the leg,
eventually decoupling the two legs at high temperatures.

We appreciate fruitful discussion with H. Tsunetsugu, J. Akimitsu, and J.
Kishine. This work is supported by the Grant in Aid of the Ministry of
Education. S. F. is supported by JSPS Research Fellowship for Young
Scientists.


\end{document}